\documentclass{article}
\pdfoutput=1

\usepackage{arxiv}

\usepackage[utf8]{inputenc} 
\usepackage[T1]{fontenc}    
\usepackage{hyperref}       
\usepackage{url}            
\usepackage{booktabs}       
\usepackage{amsfonts}       
\usepackage{nicefrac}       
\usepackage{microtype}      
\usepackage{graphicx}
\usepackage{natbib}
\usepackage{doi}

\title{CONECT4: Desarrollo de componentes basados en Realidad Mixta, Realidad Virtual Y Conocimiento Experto para generación de entornos de aprendizaje Hombre-Máquina}

\author{\href{https://orcid.org/0000-0003-2282-1945}{\includegraphics[scale=0.06]{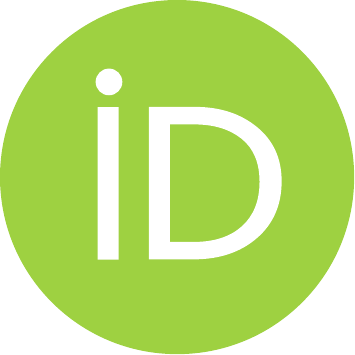}\hspace{1mm}Santiago González}\thanks{El proyecto CONECT4 está cofinanciado por el Ministerio de Industria, Comercio y Turismo a través de la línea de ayudas a las Agrupaciones Empresariales Innovadoras, en su convocatoria del 2020.} \\
	ARSoft\\
	Salamanca, Spain \\
	\texttt{santiago@arsoft-company.com} \\
	\And
	\href{https://orcid.org/0000-0002-8628-3405}{\includegraphics[scale=0.06]{orcid.pdf}\hspace{1mm}Alvaro García} \\
	Fundación Cidaut\\
	Valladolid, Spain \\
	\texttt{alvgar@cidaut.es} \\
		\And
	\href{https://orcid.org/0000-0000-0000-0000}{\includegraphics[scale=0.06]{orcid.pdf}\hspace{1mm}Ana Núñez} \\
	FACYL\\
	Valladolid, Spain \\
	\texttt{ananun@facyl.es} \\
}

\renewcommand{\shorttitle}{\textit{}}

\hypersetup{
pdftitle={CONECT4: Desarrollo de componentes basados en Realidad Mixta, Realidad Virtual Y Conocimiento Experto para generación de entornos de aprendizaje Hombre-Máquina},
pdfsubject={},
pdfauthor={},
}

\begin{document}
\maketitle

\begin{abstract}
En esta publicación se presentan los resultados del proyecto CONECT4, que aborda la investigación y desarrollo de nuevos métodos de comunicación no intrusivos para la generación de un ecosistema de aprendizaje hombre-máquina orientado al mantenimiento predictivo en la industria de automoción. A través del uso de tecnologías innovadoras como la Realidad Aumentada, la Realidad Virtual, el Gemelo Digital y conocimiento experto, CONECT4 implementa metodologías que permiten mejorar la eficiencia de las técnicas de formación y gestión de conocimiento en las empresas industriales. La investigación se ha apoyado en el desarrollo de contenidos y sistemas con un nivel de madurez tecnológico bajo que abordan soluciones para el sector industrial aplicadas en la formación y asistencia al operario. Los resultados han sido analizados en empresas del sector de automoción, no obstante, son exportables a cualquier otro tipo de sector industrial.

\end{abstract}


\section{Introducción}
Nos encontramos en un momento en el que la innovación en las empresas, y más concretamente, en el sector industrial, toma más importancia que nunca. Es necesario mejorar los niveles de productividad de las industrias para garantizar su continuidad en un mundo globalizado, donde, a medio plazo, grandes potencias como China y, a largo plazo, otros países emergentes como la India, pueden abarcar gran parte del mercado en términos del sector industrial. Esto deja a Europa en un punto de desventaja competitiva, reduciendo su capacidad de producción, lo cual supondría un golpe económico de gran importancia para el viejo continente. La mejora de la productividad debe estar soportada por una inversión en innovación que permita reducir costes, mejorar la calidad de los productos y apoyar a los trabajadores para mejorar su eficiencia. A pesar del incremento del uso de la automatización en las plantas industriales, el factor humano siempre será importante, ya que, al menos en las próximas décadas, no alcanzará el nivel tecnológico suficiente para sustituir completamente la labor humana.

No obstante, la integración de nuevas tecnologías de forma continuada en las plantas, y los cambios en los procesos de fabricación asociados a la mejora o sustitución de los productos, hacen necesario que los trabajadores tengan una formación continua. La irrupción de la pandemia COVID-19 ha cambiado las reglas del juego de la transformación digital. Si bien el teletrabajo y el aprendizaje a distancia se han convertido en algo habitual para millones de personas en la Unión Europea (UE), por otro lado se ha tenido que enfrentar a las limitaciones de los niveles de madurez en tecnologías digitales. Este fenómeno global ha acentuado la brecha en las habilidades digitales, a la vez que la desigualdad es todavía más patente si tenemos en cuenta que muchas personas no tienen el nivel requerido de estas habilidades o incluso están en lugares donde la digitalización está en niveles muy bajos.
La pandemia también ha tenido un impacto significativo en las oportunidades profesionales de muchas personas en la UE. El PIB de la UE ha sufrido una reducción del 6,6\% en 2020 mientras el desempleo se dispara hasta el 8,3\%. Este desempleo es incluso peor en los países más afectados. Además, algunos sectores específicos han experimentado una caída de más del 70\% en la facturación en el pico de mayor contagio. Para hacer frente a este desafío e impulsar la transformación digital, la UE debe abordar un cambio de paradigma en las competencias digitales. Para que las empresas de todos los tamaños, en particular las pequeñas y medianas empresas (PYMES), puedan crecer y sean más competitivas, es necesario mantenerse al día en las competencias digitales. La productividad y la innovación deben ser un objetivo sostenible a largo plazo, donde las personas dispongan de las habilidades adecuadas para trabajar de manera más eficaz aprovechando las tecnologías avanzadas. Con el aprendizaje permanente, los desajustes en el mercado laboral se reducen y se establecen las bases para la investigación y el desarrollo (I+D).

Sin embargo, según el último Innobarómetro Europeo\footnote{https://op.europa.eu/es/publication-detail/-/publication/69e52157-2ba9-11e6-b616-01aa75ed71a1}, sólo una de cada diez empresas de fabricación ha utilizado tecnologías de fabricación avanzadas (17\%), tecnologías de fabricación sostenibles (16\%) y tecnologías de fabricación inteligentes habilitadas por tecnologías de la información (11\%). La mayoría de las empresas de fabricación no han utilizado ninguna de estas tecnologías (66\%) y no tienen previsto hacerlo en un horizonte de 12 meses.
Podemos encontrar diferentes estudios para la aplicación de las tecnologías de Realidad Aumentada (RA) y Realidad Virtual (RV) en las industrias\footnote{https://www.capgemini.com/wp-content/uploads/2018/09/AR-VR-in-Operations1.pdf} \footnote{https://www.aragon.es/documents/20127/674325/Estado\%20del\%20arte\%20de\%20Realidad\%20Aumentada.pdf/f51f996d-eca5-5de4-6d07-8324ee629902}  \citep{Choi2015}, muchos de ellos asociados a la mejora de las técnicas formativas \citep{Alexander2016} \citep{Bacca2014} \citep{DiSerio2013}. 

El proyecto CONECT4 se centra en el estudio de casos concretos de sistemas y contenidos digitales basados en el concepto de Gemelo Digital (GD) y su utilización junto a herramientas de visualización avanzada como la RA y la RV, para obtener una valoración real y objetiva de escenarios de aprendizaje junto a actores clave del proceso de producción. El objetivo final es crear una serie de recomendaciones para el diseño de sistemas de comunicación hombre-máquina basadas en casos de aplicación dentro del sector de automoción que permitan la adopción de sistemas tecnológicos que realmente produzcan un impacto positivo en los resultados de las empresas. Estas recomendaciones o “manuales de buenas prácticas” permiten mejorar el diseño de productos tecnológicos orientados a la industria, gracias a la propia información que los stakeholders han podido aportar, mediante el análisis previo de los contenidos y sistemas generados en esta investigación.

De acuerdo a este planteamiento, la organización del resto del artículo será la siguiente. En la sección 2 se incluye un análisis de la situación actual de las tecnologías. En la sección 3 se describe la metodología de aprendizaje basada en la colaboración hombre-máquina. A continuación se aborda la creación de componentes y prototipos en la sección 4. Por último la sección 5 presenta los resultados y la sección 6 las conclusiones.

\section{Revisión de la tecnología}
\subsection{Estado actual de la implantación del Gemelo Digital en la Industria}
Es un hecho que, el incremento en los despliegues de sensores HMI (Human Machine Interface) avanzados para registrar información en ecosistemas de colaboración hombre-máquina, y en las tendencias tecnológicas de actualidad como la Inteligencia Artificial (IA), en especial el aprendizaje automático, están potenciando el empleo de GDs. Esta tecnología habilitadora de la Industria 4.0 (I40) permite crear una representación virtual de un producto o servicio del mundo real de forma conectada \citep{Tao2019}. La posibilidad de representar procesos de forma virtual con sus indicadores físicos asociados en tiempo real, introduce modelos digitales no intrusivos de casos de uso industrial. Este concepto se convierte en una valiosa herramienta para la toma de decisiones gracias a la realimentación de los datos en los procesos de aprendizaje y generación de conocimiento, mediante la aplicación de tecnologías de análisis de datos, simulación, visualización asistida, e IA. La investigación de metodologías de comunicación no intrusivas apoyadas en las Tecnologías de la Información y las Comunicaciones (TICs) para la fábrica digital y conectada, presenta un nuevo paradigma de unión entre la planta productiva y las nuevas tecnologías digitales, a su vez representado por los procesos y los sistemas de información (convergencia de tecnologías operacionales y tecnologías de la información) \citep{Odonovan2016}. Esta visión representada por el GD permite la generación de ecosistemas de aprendizaje para la toma de decisiones de forma más rápida y ágil: (i) Teniendo información del ciclo de vida de los diferentes procesos y sus interacciones con la posibilidad de generar conocimiento a partir de sus estados e indicadores en tiempo real \citep{Macchi2018}; (ii) Analizando decisiones estratégicas de forma centralizada y efectiva sin interferir o paralizar los procesos productivos, anticipándose con un enfoque predictivo, no reactivo \citep{Erikstad2017}, y (iii) Incorporando modelos de datos distribuidos a través de diferentes puntos de la cadena de proceso y de suministro \citep{Banerjee2017}. 
A través de este concepto de GD se adquiere una visión más detallada acerca del comportamiento de una instalación industrial ante diversas situaciones y modelos analizados para la toma de decisiones anticipadas. Todo ello sin alterar el transcurso de la producción, con un alto grado de exactitud y sin que esto suponga un coste económico adicional a la empresa.

De acuerdo a diferentes estudios, este tipo de tecnología digital proporciona un entorno donde las empresas de todos los tamaños puedan hacer réplicas de sus activos para detectar problemas de manera temprana, conseguir reducir el tiempo de fabricación casi un 50\%, disminuir la energía consumida en un 70\%, mejorar la flexibilidad e incrementar la productividad en hasta un 20\%\footnote{https://ciudadesdelfuturo.es/que-es-el-gemelo-digital.php}, lo que determina una mayor competitividad global. Grandes referentes tecnológicos como Gartner lleva varios años situando al GD entre las diez principales tendencias tecnológicas, y firmas de análisis también referentes, como Juniper Research, pone cifra a su relevancia. En concreto, Juniper Research en su informe “Digital Twins in IoT: Market Strategies, Challenges \& Future Outlook, 2019-2023” estima que los ingresos derivados del mercado de GDs crecerán a un ritmo promedio del 35\% anual desde los 9.800 millones de dólares, hasta alcanzar los 13.000 millones de dólares en 2023. A mayores indica que, el 34\% de los ingresos en 2023 procedentes de GDs vendrán por la parte de manufactura y fabricación, hasta un total estimado de 4.500 millones de dólares. Le siguen en relevancia el sector del transporte, que facturará 2.500 millones de dólares en relación a los GDs, y energía y servicios públicos, con 1.100 millones. 

Según los informes de PricewaterhouseCoopers (PwC)\footnote{https://www.strategyand.pwc.com/gx/en/insights/industrial-internet.html}, la industria basada en datos ya está generando más de 110 mil millones de euros de ingresos adicionales en Europa. Por su parte, McKinsey \& Company\footnote{https://www.mckinsey.com/business-functions/mckinsey-digital/our-insights/digital-in-industry-from-buzzword-to-value-creation} estima que la I40 mejoraría la productividad en la profesión técnica entre un 45 y un 55\%. Trabajos recientes inciden en que uno de los principales problemas dentro de las empresas existentes está relacionado con sistemas de adquisición de datos incompletos \citep{Weyer2015}. Se subraya que la infraestructura de datos es uno de los aspectos más desafiantes para las empresas en la transición hacia la I40. Se hace evidente la necesidad de estandarizar y modular la infraestructura de datos para generar conocimiento e inteligencia dentro de los sistemas de producción. Por lo tanto, es posible afirmar que, existe a día de hoy una brecha de investigación para el desarrollo de un enfoque eficiente de "talla única" para el análisis de procesos impulsado por datos y la adquisición de datos. Otro aspecto al que debemos hacer referencia es la relación entre los GDs y los sistemas ciber-físicos. Desde que \citep{Grieves2015} presentara el primer concepto de GD para comprender mejor la producción y el diseño utilizando una réplica de fábrica virtual, hasta que \citep{Tao2018} presentara una guía concisa para el diseño de productos impulsados por GDs, es un hecho que implementar GDs dentro de las infraestructuras existentes no es una tarea fácil; no existiendo a día de hoy una solución estandarizada. Investigadores como \citep{UHLEMANN2017335} afirman que los desafíos relacionados con los GDs incluyen aspectos como: dificultades en la adquisición de datos en tiempo real, en la definición de los requisitos para los sistemas e infraestructura de información, implementación incorrecta, estandarización de los sistemas de adquisición de datos, costos de inversión, integridad débil entre el mundo físico y cibernético y la seguridad de los datos.

A pesar de todas estas dificultades, la infraestructura de GDs se ha establecido en varias empresas líderes del sector, como General Electric, PTC, Siemens, Oracle, ANSYS, Dassault, SAP y Altair \citep{QI2018237}. Sin embargo, el GD es todavía un concepto de desarrollo \citep{Tao2019} que, además presenta diversas barreras tecnológicas para su adopción en el tejido industrial. Por un lado, existe una dificultad técnica para monitorizar masivamente y digitalizar procesos en la industria \citep{Glass2018}, con gran variedad de equipos, sistemas heredados aislados \citep{Chesworth2018}, buses de campo, protocolos propietarios, así como una estricta arquitectura de integración y automatización industrial \citep{Helu2017}. Por otro lado, los sistemas utilizados actualmente en la industria son incapaces de almacenar y tratar los volúmenes de datos necesarios para crear y evolucionar los GDs, que realmente representen el comportamiento de los elementos físicos, y no sólo sus características y su estado \citep{Tao2018b}.

\subsection{Estado actual de implantación de la Realidad Aumentada y Realidad Virtual en la industria}
Actualmente es un hecho que el COVID-19 ha acelerado los planes de transición digital de muchas empresas, como destaca la consultora tecnológica Deloitte . No obstante, por otro lado, se da un efecto contrario donde otras empresas han parado sus inversiones en este campo, buscando un ahorro mediante una reducción en la inversión en innovación. Esto, en este tipo de circunstancias, representa un error de base, ya que el beneficio de esta inversión permite aumentar la productividad y por tanto maximizar los beneficios en el corto-medio plazo. Según los informes de PricewaterhouseCoopers (PwC) la RA y la RV tendrán un impacto de hasta 1,5 billones de dólares en la economía mundial en 2030. La RA tiene multitud de aplicaciones en la Industria, desde la formación de trabajadores, hasta la asistencia remota o la mejora de la visualización de datos complejos en tiempo real. Todas estas posibilidades están empezando a ser aprovechadas por las industrias, consiguiendo mejorar los tiempos y en definitiva la productividad de las empresas. Por otro lado, la RV permite una inmersión completa en un mundo virtual, siendo posible la interacción para realizar simulaciones de procesos con el objetivo de conseguir una formación autónoma y deslocalizada. En cualquier caso, la realidad que nos encontramos, es que las diferentes empresas han declarado la dificultad que tienen a la hora de encontrar personal cualificado en nuevas tecnologías para incorporar en sus equipos de trabajo. De hecho, la Comisión Europea ha establecido indicadores ambiciosos de capacitación, reevaluación y mejora de las competencias, vinculados con la recuperación económica a la pandemia, que incluyen:
\begin{itemize}
    \item Para 2025: 120 millones de adultos en la UE deberían participar en actividades formativas cada año. Esto corresponde al 50\% de la población adulta y alrededor de 540 millones de actividades de formación para este colectivo durante el quinquenio.
    \item Para 2025: 2 millones de personas que buscan empleo o uno de cada cinco deberían tener una experiencia de aprendizaje reciente. 
    \item Para 2025: 230 millones de adultos deberían tener al menos competencias digitales básicas, lo que cubre el 70\% de la población adulta de la UE.
\end{itemize}

\begin{figure}[ht]
    \centering
    \includegraphics[width=0.7\textwidth]{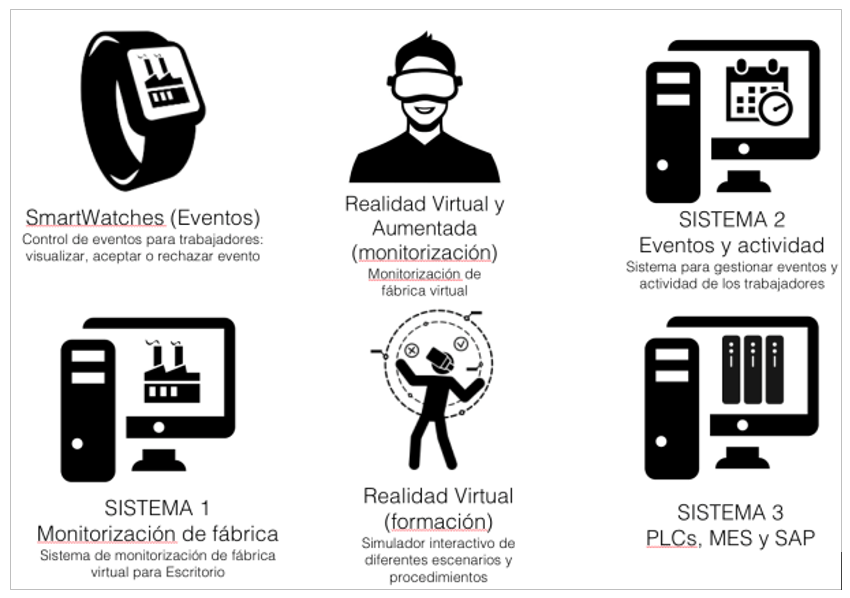}
    \caption{Nuevas aplicaciones y plataformas que permiten tener acceso a la información de las fábricas en tiempo real (ARSoft)}
    \label{fig1}
\end{figure}

Según la propia comisión, para el cumplimiento de estos objetivos se realizará una inversión adicional estimada en 48 billones de Euros anuales. Todos estos datos reflejan la necesidad actual de disponer de herramientas formativas de calidad, siendo la RA y RV las tecnologías idóneas para llevarlas a cabo. 
Actualmente tan sólo una cantidad minoritaria de empresas industriales, o entidades educativas como Universidades, está utilizando estas tecnologías como herramientas formativas (ver Figura \ref{fig1}). 

Lo más habitual es encontrar empresas que sí han realizado algún tipo de desarrollo con RA o RV, no obstante lo han hecho dentro de una fase piloto de testeo de tecnología. Atribuimos esta situación a que entre 2013 y 2017 muchas empresas realizaron inversiones en estas tecnologías, que finalmente resultaron no productivas. Esto fue debido al estado de la tecnología en ese momento, que no estaba lo suficientemente madura para que las empresas pudieran obtener rentabilidad de sus inversiones.
No obstante, a día de hoy la tecnología ha avanzado mucho, disponiendo de dispositivos con gran capacidad de procesamiento y renderizado capaces de ofrecer experiencias, tanto de RA como de RV, de gran calidad. Además, el coste de estos dispositivos se ha reducido drásticamente. Estos avances han permitido que algunas empresas sí estén apostando por estas tecnologías, aplicando la RA para los procesos de mantenimiento y la RV para los procesos formativos. En el sector industrial, estas inversiones se hacen más evidentes en sectores como el naval o aeroespacial, mientras que en el sector académico muchas universidades disponen ya de simuladores de RV para formar a sus estudiantes. Estos avances permiten proyecciones de mejora de inversión en estas tecnologías como los siguientes:
\begin{itemize}
    \item Sector de la automoción: En 2019 las inversiones en Realidad Virtual eran de unos 760 millones de dólares, proyectando una tasa de crecimiento anual compuesto (CAGR) superior al 45\%.
    \item Sector energético: Según el Foro Económico Mundial la tecnología digital podría alcanzar un valor de 1 trillón de dólares en los próximos 10 años.
    \item Sector aeroespacial y de defensa: En 2018 el mercado global de Realidad Virtual era de aproximadamente 400 millones de dólares, proyectando un CAGR del 38\%.
    \item Sector sanitario: La Realidad Aumentada alcanzaba en 2019 una inversión de 780 millones de dólares en 2019, esperando para el año 2025 una inversión total de más de 4.000 millones de dólares.
    \item Sector de la Educación Superior: El crecimiento en este sector se espera con un CAGR de 16,2\%, con un mercado de 9,3 billones de dólares en 2018 que será de 19,6 billones en 2023.
\end{itemize}

Como conclusión, podemos afirmar que los nuevos ecosistemas de colaboración hacen que las personas se conviertan en una parte más del GD para entender qué está pasando. Se incorporan capacidades de visualización por medio de dispositivos conectados como smartphones, tablets o gafas de RA, lo que permite conocer, analizar e interaccionar de forma adaptativa partiendo de los datos que se están registrando en tiempo real. Podemos dar soporte a procesos de mantenimiento para saber cuándo una máquina va a fallar con solo mirarla, la temperatura que se está registrando, su consumo, o su curva de funcionamiento comparada con la referencia ideal para detectar comportamientos erróneos o simplemente mejorables. Por otro lado, la RV permite visualizar y manejar información de una forma completamente diferente a las interfaces habituales.

\section{Metodología}
El potencial de la transformación digital de las empresas productivas necesita de la colaboración hombre-máquina, apoyada en la formación de nuevos perfiles técnicos y en un aprendizaje de las tecnologías habilitadoras aplicadas a los procesos de fabricación. El rápido movimiento hacia la digitalización, derivado de los avances en las nuevas tecnologías habilitadoras de la I40, supone un gran avance con la introducción de sistemas que ayuden a la toma de decisiones de forma más rápida y ágil. La conectividad del Internet Industrial de las Cosas (IIoT) unido al conocimiento experto con técnicas de Machine Learning (ML), habilita la gestión, trazabilidad y mantenimiento de procesos industriales. La convergencia entre el mundo físico y virtual que aporta el concepto de GD, incorpora a su vez la visualización y formación en planta con la RA y la RV, poniendo la tecnología al alcance de las diferentes personas que trabajan junto a las máquinas.

\begin{figure}[ht]
    \centering
    \includegraphics[width=0.8\textwidth]{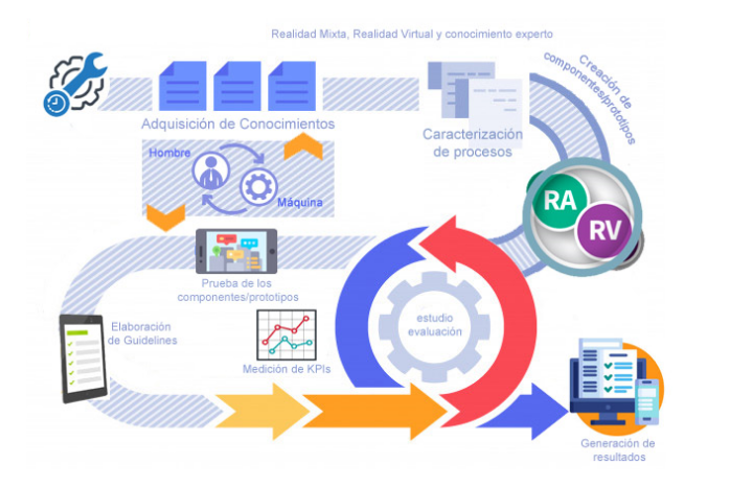}
    \caption{Generación de modelos de aprendizaje basados en la colaboración hombre-máquina}
    \label{fig2}
\end{figure}

Actualmente, las capacidades de la IA están más allá de la capacidad humana en muchos dominios. Por lo tanto, un enfoque bien encaminado, se sustenta en la construcción de un marco de aprendizaje capaz de explotar conjuntamente la inteligencia humana y la inteligencia del sistema. Hace unos años, los sistemas de coaching digital \citep{Carlsson2018} comenzaron como una respuesta a la demanda de operadores humanos capaces de gestionar sistemas automatizados avanzados capaces de monitorizar y controlar complejos procesos y sistemas industriales complejos. Hoy en día, el uso más común de la IA dentro de la industria de fabricación está diseñado para ser considerado unidireccional. Los seres humanos aprovechan la información proporcionada por los diferentes sistemas, pero los sistemas no pueden aprovechar el conocimiento y el “know how” del ser humano. La caracterización de métodos de aprendizaje en procesos colaborativos, y la aplicación de técnicas de análisis inteligente sobre los conjuntos de datos almacenados del entorno productivo, permite a su vez definir modelos de comportamiento en torno a nuevas tecnologías habilitadoras digitales (THDs) como el GD, la RA, la RV. En el proyecto CONECT4 se presenta cómo estas tecnologías, apoyadas en la investigación y creación de nuevos modelos de IA no intrusivos, pueden utilizar como base patrones de conocimiento dirigidos por personas. Esta interacción sirve para descubrir las directrices de aprendizaje, a partir de resultados experimentales y la utilización de bancos de pruebas (ver Figura \ref{fig2}). 

Se plantea por tanto, la reproducción de un entorno de interacción que permita:
\begin{itemize}
    \item Mejorar los procesos de entrenamiento por medio del uso de la RV y simuladores interactivos, creados con una plataforma que permita la instalación de estos contenidos a gran escala.
    \item Mejorar las contenidos de procedimientos para mantenimiento por medio de la RA, creados con una plataforma que permita la instalación de estos contenidos a gran escala.
    \item Caracterizar y explicar un conjunto de problemas asociados a procesos industriales donde intervienen frecuentemente procedimientos que dependen de personas y máquinas.
    \item Utilizar medios cognitivos a partir de técnicas de aprendizaje automático semi-supervisado para mejorar la toma de decisiones, con la comprensión de cómo las personas por sí mismas pueden mejorar las capacidades de las máquinas de manera más efectiva, incrementando la flexibilidad y personalización de los procesos a pie de planta.
\end{itemize}

De acuerdo al esquema de la Figura \ref{fig2}, que presenta el concepto general basado en la colaboración hombre-máquina, podemos caracterizar esta metodología de aprendizaje colaborativo atendiendo a 7 fases:

\paragraph{Fase 1. Adquisición de conocimientos.}
El objetivo principal de esta primera fase consiste en comprender la interacción de los sistemas y personas que trabajan en el entorno productivo. Para ello, se considera el estudio de entornos industriales heterogéneos formados por diferentes tamaños de empresas, donde entender en una muestra representativa los aspectos y particularidades para la colaboración efectiva hombre-máquina. La información analizada se emplea para construir los esquemas de conocimiento, de forma que sea posible su correcta organización y estructuración, y por tanto útil, a la hora de generar los modelos de comportamiento hombre-máquina. A partir de esta información, se determinan las fuentes de conocimiento suficientes para completar la tarea (expertos, operarios, bibliografía especializada, manuales, etc.). Además se tienen en cuenta distintos esquemas de razonamiento considerados para modelizar las diferentes necesidades de los ecosistemas industriales identificados. Por otro lado, se analiza la información disponible, así como su fiabilidad y coherencia para desarrollar un modelo formal del problema con el que el sistema experto pueda razonar.
\paragraph{Fase 2. Caracterización de los procesos.}
El objetivo de esta segunda fase consiste en la caracterización de entornos de aprendizaje basados en el concepto de GD. Para ello se consideran los  patrones de conocimiento adquiridos de acuerdo a la visión holística integrada por personas - máquinas - procesos. Las tareas o actividades que conforman el proceso estudiado, la secuencia individual de operaciones que se suceden, así como las personas que intervienen son identificadas. A partir de esta información se analizan los pasos necesarios para convertir los procesos actuales en procesos más ágiles con la posibilidad de incluir automatización de tareas. Se identifican posibles riesgos y se descubren qué pasos son susceptibles de mejora. A su vez, se incluyen nuevos controles, registros de control o indicadores del proceso. Como resultado, se consigue una representación virtual del proceso de producción de modo que se pueda examinar su diseño, probar potenciales cambios y detectar errores de forma experimental antes de llevarlo al entorno real productivo.
\paragraph{Fase 3. Creación de componentes/prototipos.}
El objetivo de esta tercera fase consiste en la elaboración de prototipos en entorno de laboratorio para simular la interacción industrial hombre–máquina mediante la utilización conjunta de Realidad Mixta e IA. Adicionalmente se realiza un análisis de usabilidad. El empleo de estas tecnologías habilitadoras aplicadas a la generación de conocimiento permite:
\begin{itemize}
    \item Obtener modelos predictivos a través de técnicas de análisis de datos con algoritmos de ML orientados a predecir resultados futuros basados en datos históricos y huellas digitales características. Se consigue así que los datos capturados en planta obtengan valor y sean más transparentes a la hora de aportar conocimiento.
    \item Elaborar visores aumentados que proporcionan herramientas digitales a los trabajadores para mostrar toda la información de un activo, incluyendo manuales de funcionamiento, modelos 3D con todos los componentes que lo integran y valores de estado en base a la información registrada y la interpretación que los modelos analíticos hayan proporcionado.
    \item Implementar simuladores de RV a través de dispositivos inteligentes como gafas y tabletas. Permiten al usuario interactuar con los objetos que conforman el proceso industrial.
    \item Crear visualizaciones de GD para tener una representación virtual del proceso físico que permita comprender y predecir las condiciones deseadas de rendimiento de un proceso a partir de la simulación y optimización.
\end{itemize}
\paragraph{Fase 4. Estudio de evaluación.}
El objetivo de esta cuarta fase consiste en validar las diferentes técnicas de aprendizaje en un entorno experimental con los datos facilitados por los trabajadores de las diferentes empresas que colaboran en el proyecto. En primer lugar es necesario verificar que los modelos virtuales de procesos generados en un entorno de laboratorio se ajustan a los escenarios elegidos para mejorar la comunicación hombre-máquina. Las técnicas de evaluación nos permiten comprobar en qué medida los resultados obtenidos en las simulaciones realizadas coinciden con los objetivos planteados para al diagnosis, permitiendo la mejora en la adquisición de conocimientos y retroalimentar la caracterización de los procesos para generar prototipos mejorados.
\paragraph{Fase 5. Prueba de los componentes/prototipos.}
El objetivo de esta quinta fase consiste en llevar a cabo la ejecución y aplicación de los prototipos implementados en entorno de laboratorio para medir diferentes indicadores que permitan obtener los resultados de la investigación y demostrar cómo siguiendo unas pautas de implantación tecnológica se puede conseguir la generación de conocimiento industrial orientado al mantenimiento predictivo. Para ello es necesario realizar un proceso de comparación entre situaciones reales y situaciones simuladas que determinen si los requisitos de usabilidad, inteligibilidad y los resultados obtenidos son válidos o se obtienen desviaciones. Esta prueba se realiza de forma iterativa de manera que sea posible introducir las mejoras necesarias para conseguir que los resultados cumplan con los requisitos para la gestión efectiva del conocimiento del proceso.
\paragraph{Fase 6. Medición de KPIs.}
El objetivo de esta sexta fase consiste en la medición y representación de los indicadores clave que se utilizan para valorar el grado de cumplimiento de los objetivos previamente establecidos. Estos KPIs o indicadores clave se utilizan para determinar el estado funcional del prototipo y permiten definir una línea de acción futura, es decir, expresan las variables que ayudan a la toma de decisiones. Por tanto, para conocer el grado de adecuación de las técnicas a los resultados esperados, se realizan mediciones con indicadores del proceso a partir de los contenidos y métodos aumentados en la muestra seleccionada en las diferentes empresas objeto del estudio, con distintos perfiles de trabajadores de planta. 
\paragraph{Fase 7. Generación de resultados y guidelines.}
El objetivo de esta séptima y última fase consiste en la generación de la documentación y la preparación de contenidos para la difusión de los resultados de la investigación. Con esta documentación es posible disponer de guidelines de implantación tecnológica y generación de productos de Realidad Mixta y Conocimiento Experto para la mejora de los procesos de soporte, formación e integración con GD en procesos industriales de mantenimiento.

Es importante destacar que la adopción de procesos guiados semi-autónomos se encuentra todavía en una etapa inicial en la industria, donde los desafíos específicos se convierten en una barrera para su introducción: (i) Por la larga vida útil de los activos o infraestructuras caracterizadas por procedimientos de operación cerrados;  (ii) Debido a objetivos comerciales basados en la reducción de costes y de los recursos para mejorar continuamente la calidad y la productividad; (iii) Por los orígenes de datos de fuentes heterogéneas y con acceso propietario (no abierto) a interfaces de comunicación; (iv) Debido a requisitos cada vez más altos en cuanto a confiabilidad, seguridad, protección y cuestiones éticas; (v) Por las prácticas sociales habituadas a afrontar sucesos inesperados utilizando enfoques tradicionales heredados; y (vi) Por la falta de conocimiento experto y de las habilidades tecnológicas requeridas para abordar procesos de capacitación. Por eso es importante la aplicación y evaluación práctica de las tecnologías y su interacción con personas como se verá en la siguiente sección.

\section{Creación de componentes y prototipos}
El estado del arte actual en la realización de procesos formativos en la industria no ha cambiado mucho en las últimas décadas, disponiendo de una figura de formador, el cuál se encarga de enseñar a un grupo de empleados cómo llevar a cabo determinadas tareas. En algunas empresas, gracias  la incorporación de plataformas de eLearning como Moodle, los empleados pueden acceder a cursos formativos. Sin embargo esto suele llevarse a cabo para una formación teórica, y no tanto para formación práctica, que esté asociada al mantenimiento de máquinas o la realización de procesos de fabricación.

La utilización de soluciones basadas en la RV, permite a los usuarios formarse de manera autónoma y completamente práctica, pudiendo incluso recibir una evaluación automática, de forma que el propio sistema es capaz de certificar que un empleado está perfectamente formado en un proceso concreto.
El proyecto CONECT4 propone la creación de un marco de aprendizaje colaborativo y conectado, de forma no intrusiva, donde se aprovecha el conocimiento en procesos de asistencia a los trabajadores de la industria de forma predictiva. Herramientas como la Realidad Mixta y las interfaces hombre-máquina están listas para generar nuevos paradigmas y procesos de asistencia guiada en la industria. Sin embargo, tradicionalmente los trabajadores de un sistema necesitan pasar por un período de aprendizaje así como por una orientación y un asesoramiento supervisado (en tiempo real, con datos e información reales). Esta información debe adaptarse a sus capacidades, habilidades y al conocimiento previo de cada individuo para permitir que los trabajadores progresen en el uso del sistema, por ejemplo a través de consejos que puedan comprender y usar, etc. Por esta razón, la creación de diferentes prototipos tecnológicos permiten evaluar métodos de aprendizaje y de interacción hombre-máquina en entornos de mantenimiento industrial.

\subsection{Contenidos de Realidad Aumentada y Realidad Virtual para la mejora de los procesos de mantenimiento y procesos formativos}
Dentro de esta fase de creación de contenidos utilizados en métodos de aprendizaje, se ha seguido la metodología presentada en la sección 3 con la implementación de un escenario en el que trabajar con simuladores de RV, a partir de los siguientes pasos:
\begin{itemize}
\item Obtención de información relacionada con el proceso industrial, donde se utiliza como punto de partida un procedimiento técnico de actuación.
\item Obtención de recursos asociados, como son imágenes, videos o documentos procedentes del proceso industrial, que puedan ser incorporados en el simulador de RV.
\item Preparación del contenido, donde se crea un simulador de RV en base a la información proporcionada por la empresa industrial en los pasos anteriores. En este punto, se puede planificar la incorporación de datos procedentes de sensores o sistemas de monitorización conectados en plataformas cloud, o incluso datos proporcionados por una sistema experto (como por ejemplo la generación de una alerta asociada a la condición del proceso).

\begin{figure}[ht]
    \centering
    \includegraphics[width=0.8\textwidth]{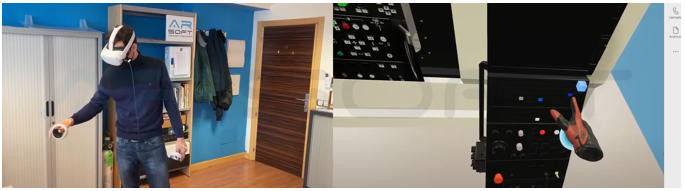}
    \caption{Demostración de tecnologías de Realidad Virtual para entornos de fabricación}
    \label{fig3}
\end{figure}

\item Medición de KPIs asociados al proceso industrial a partir de una prueba sobre los contenidos con diferentes perfiles de empleados, para conocer el grado de impacto de estos contenidos en su actividad. Algunas de las métricas utilizadas con el tiempo de formación invertido o el número de errores cometidos (Figura \ref{fig4}).
\item Realización de una encuesta, donde los usuarios de la aplicación facilitan indicadores para conocer su grado de satisfacción y complementar la información obtenida en el paso anterior.
\end{itemize}

El sistema presenta diferentes modos de ejecución de los contenidos formativos: (i) Modo Demo, el usuario tan sólo debe ver las instrucciones a llevar a cabo. No dispone de opción de interacción. (ii) Modo Guiado, el usuario debe llevar a cabo por sí mismo las diferentes tareas del proceso, pero en todo momento el sistema le guía y le proporciona información sobre la acción que debe realizar. (iii) Modo Manual, el usuario interactúa con el sistema, pero no dispone de ninguna ayuda. (iv) Modo Evaluación, el usuario no dispone de ninguna ayuda, y al final del proceso recibe una evaluación automática que determina si ha finalizado su formación.

\begin{figure}[ht]
    \centering
    \includegraphics[width=0.7\textwidth]{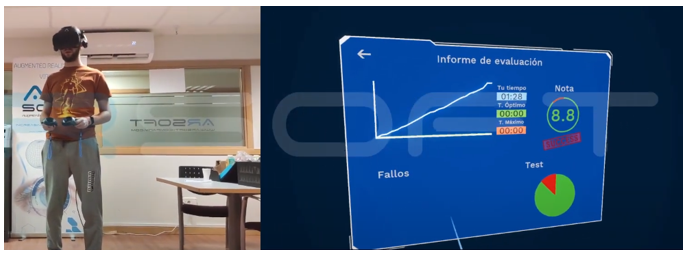}
    \caption{El sistema de RV es capaz de generar una evaluación automática del usuario en un proceso concreto en base a diferentes parámetros}
    \label{fig4}
\end{figure}

\subsection{Sistemas de aprendizaje e interfaces de acceso a información del Gemelo Digital para mejora de la eficiencia en la comunicación Hombre-Máquina}

Utilizando la misma infraestructura de creación de contenidos digitales, apoyada en herramientas conectadas de monitorización y de interacción con los trabajadores a través de plataformas cloud, se han evaluado métodos de aprendizaje que se apoyan en la visualización de datos aumentados a través de la integración del concepto del GD.
Este nuevo enfoque de comunicación hombre-máquina se utiliza para validar la introducción de las tecnologías digitales dentro del concepto de “fábrica conectada” o “fábrica inteligente” para la I40.

\begin{figure}[ht]
    \centering
    \includegraphics[width=0.4\textwidth]{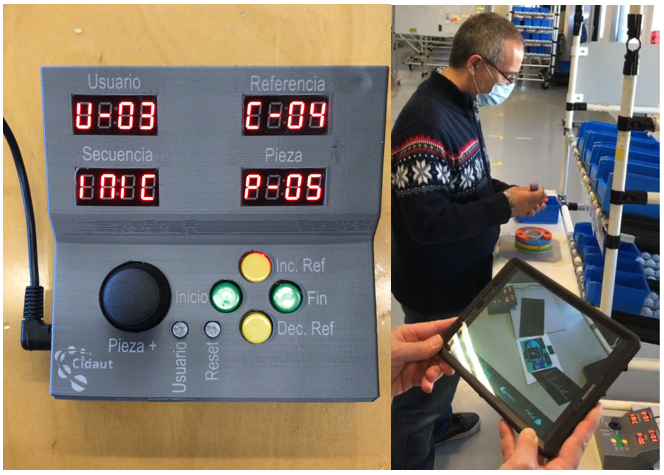}
    \caption{Digitalización del proceso formativo (Escuela Lean)  utilizando herramientas interactivas y RA para visualización}
    \label{fig5}
\end{figure}

Gracias a la colaboración de personal de la Escuela Lean de la Universidad de Valladolid, se ha implementado una prueba de concepto (ver Figura \ref{fig5}) que permite de forma sencilla medir y monitorizar con indicadores cuantitativos (número de operaciones, tiempo empleado) y con indicadores cualitativos (progreso, dificultad), el nivel de mejora que se obtiene a través de procesos digitalizados durante la formación de personas que trabajan en entornos industriales. La Escuela Lean dispone de diferentes productos \citep{Gento2020} que permiten caracterizar y adaptarse a un entorno real de la fábrica, junto a las necesidades del proceso donde participan varios perfiles de trabajadores. La prueba de concepto consta de un dispositivo HMI conectado en tiempo real para el envío de la información y de un software de visualización instalado en una Tablet. Este software de RA se utiliza para que el formador reciba la información de seguimiento del proceso de forma instantánea, asociada al proceso que se está realizando con un código QR, completando la integración del GD.

El sistema permite introducir varios modos de ejecución de los contenidos formativos, siendo configurable para un conjunto variable de puestos Lean. El usuario tan sólo debe fijar la secuencia de inicio, comenzar con el proceso formativo que le ha indicado el instructor, interaccionando con un pulsador cada vez que completa una operación concreta. Para acabar unicamente debe fijar la secuencia de finalización para comunicar la detención del proceso. 

Este entorno de componentes digitalizados se ha mostrado de utilidad para abordar metodologías de formación en procesos de mejora continua (fabricación, mantenimiento) sobre escenarios industriales. Permite analizar con los formadores la aplicación de tecnologías habilitadoras, en este caso con la digitalización de un proceso formación unido a su visualización con un GD y seguimiento mediante RA, proporcionando herramientas de mejora en los procesos de aprendizaje.

\section{Resultados}
Como análisis de los resultados de aplicabilidad de las metodologías y tecnologías presentadas en el proyecto CONECT4, se incluye una encuesta en la que han participado para su evaluación diferentes actores clave de empresas industriales del sector de automoción (incluyendo PYMES y grandes empresas). Esta encuesta presenta los siguientes objetivos:

\begin{itemize}
\item Valorar el estado del arte de estas tecnologías en las empresas industriales
\item Valorar el grado de aceptación de los contenidos digitalizados
\item Valorar la experiencia una vez ejecutados los contenidos
\item Valorar el grado de impacto de estos contenidos en la actividad de la empresa
\end{itemize}

\begin{figure}[ht]
    \centering
    \includegraphics[width=0.4\textwidth]{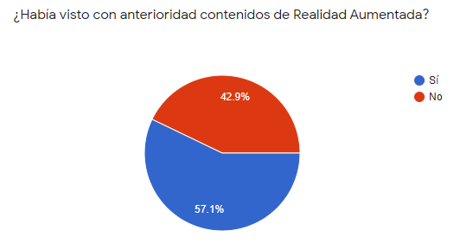}
    \caption{Nivel de conocimiento de las empresas participantes sobre aplicaciones de RA}
    \label{fig6}
\end{figure}

\begin{figure}[h!]
    \centering
    \includegraphics[width=0.4\textwidth]{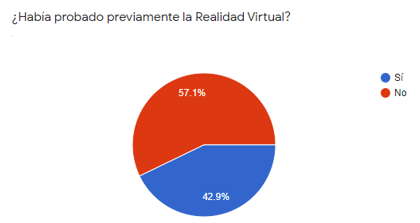}
    \caption{Nivel de conocimiento de las empresas participantes sobre aplicaciones de RV}
    \label{fig7}
\end{figure}

En primer lugar, se ha podido detectar que las tecnologías de RA, RV y GD sí se consideran para su estudio y aplicación en las diferentes empresas, estando más presente en grandes empresas que en PYMES. No obstante, este grado de implantación es aún bajo, muchas veces limitándose a acciones concretas y muchas veces con acciones enmarcadas dentro de proyectos pilotos. Aún así, sí se aprecia que las empresas están empezando a apostar más por las tecnologías a medio y largo plazo (Figura \ref{fig6}, Figura \ref{fig7}).

\begin{figure}[h!]
    \centering
    \includegraphics[width=0.5\textwidth]{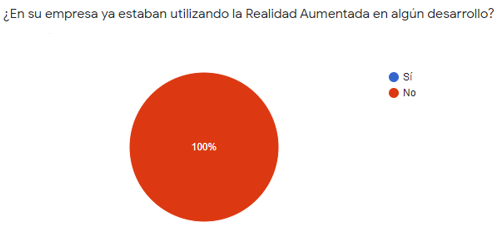}
    \caption{Nivel de uso de aplicaciones de RA por parte de las empresas participantes}
    \label{fig8}
\end{figure}

\begin{figure}[h!]
    \centering
    \includegraphics[width=0.5\textwidth]{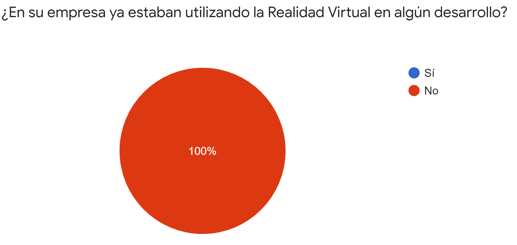}
    \caption{Nivel de uso de aplicaciones de RV por parte de las empresas  participantes}
    \label{fig9}
\end{figure}

Aunque las siguientes preguntas han sido contestadas negativamente por el 100\% de los encuestados, hay que recalcar que las empresas sí habían realizado acciones en torno a las tecnologías de RA y RV, no obstante las preguntas estaban dirigidas a la implantación de las tecnologías fuera de casos concretos o pilotos, sino como tecnología integrada en los procesos de la empresa (Figura \ref{fig8}, Figura \ref{fig9}).

\begin{figure}[h!]
    \centering
    \includegraphics[width=0.7\textwidth]{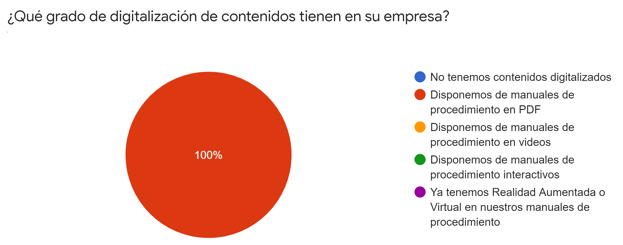}
    \caption{Grado de digitalización de contenidos en PDF de las empresas participantes}
    \label{fig10}
\end{figure}

La Figura \ref{fig10} muestra que el 100\% de las empresas consultadas utilizan los PDF o documentos digitales como metodología de acceso a la información de los procesos de mantenimiento. Por tanto el grado de digitalización de las empresas en cuanto a formación y mantenimiento es bajo o muy bajo.

\begin{figure}[h!]
    \centering
    \includegraphics[width=0.7\textwidth]{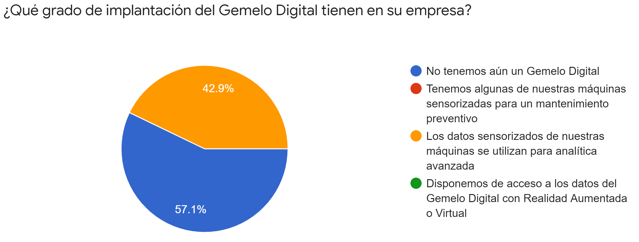}
    \caption{Grado de implantación de la tecnología del GD por parte de las empresas participantes}
    \label{fig11}
\end{figure}

Respecto al grado de implantación de los GDs (Figura \ref{fig11}) , aquí se encuentra una diferencia importante en función del tamaño de las empresas encuestadas, donde las empresas grandes sí disponen de un GD implantado, mientras que las empresas más pequeñas cuentan con máquina sensorizadas pero no disponen de una plataforma de GD implantada.

\begin{figure}[ht]
    \centering
    \includegraphics[width=0.9\textwidth]{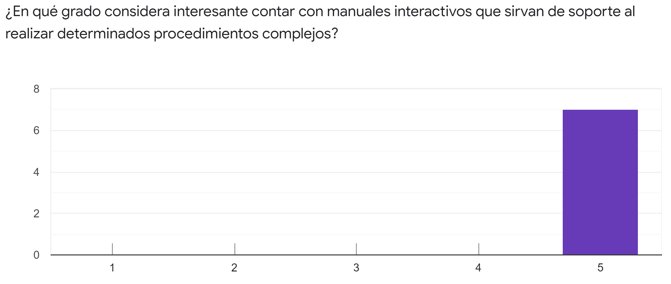}
    \caption{Grado de interés acerca del uso de manuales interactivos por parte de las empresas participantes}
    \label{fig12}
\end{figure}

\begin{figure}[h!]
    \centering
    \includegraphics[width=0.9\textwidth]{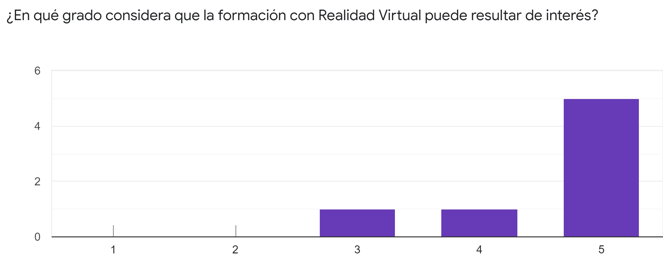}
    \caption{Grado de interés acerca de la formación con RV por parte de las empresas participantes}
    \label{fig13}
\end{figure}

\begin{figure}[h!]
    \centering
    \includegraphics[width=0.9\textwidth]{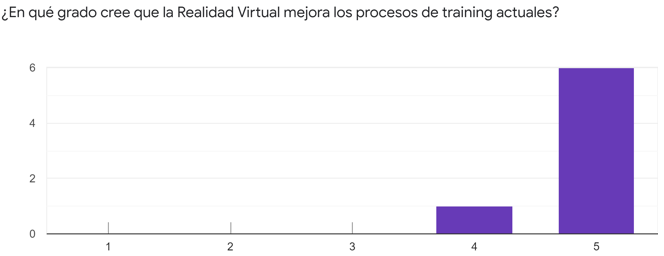}
    \caption{Grado de interés acerca de la mejora de procesos de formación con RV por parte de las empresas participantes}
    \label{fig14}
\end{figure}

Respecto al uso de tecnología 3D interactiva, con RA, RV, o simplemente 3D (Figura \ref{fig12}, Figura \ref{fig13}, Figura \ref{fig14}), las respuestas de los usuarios muestran un gran interés en todos los sentidos.

\begin{figure}[h!]
    \centering
    \includegraphics[width=0.9\textwidth]{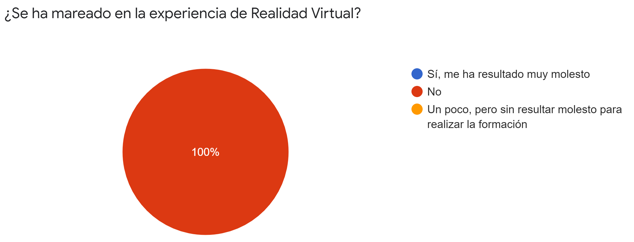}
    \caption{Usabilidad de la tecnología de RV}
    \label{fig15}
\end{figure}

Se ha valorado la experiencia de usuario en cuanto a los contenidos de RV respecto a la sensación de mareo (Figura \ref{fig15}), ya que los dispositivos de RV hace unos años sí provocaban esta sensación. No obstante, con los dispositivos actuales se demuestra que esta sensación desaparece en prácticamente el 100\% de los casos, habiéndose empleado para esta prueba concreta el dispositivos Oculus Quest 2, que se encuentra entre los dispositivos actuales de gama media-alta.

\begin{figure}[h!]
    \centering
    \includegraphics[width=0.9\textwidth]{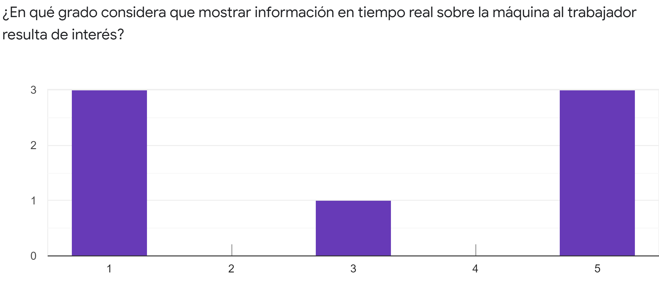}
    \caption{Grado de interés acerca del uso de información en tiempo real por parte de las empresas participantes}
    \label{fig16}
\end{figure}

En relación a las ventajas que los usuarios encuentran respecto a la visualización de información en tiempo real de la maquinaria (Figura \ref{fig16}), nos hemos encontrado opiniones muy diversas, puesto que algunos encuestados no lo encontraban nada interesante, mientras que otros se posicionaron en el extremo contrario, mostrando un gran interés. Esto se debe a que estas ventajas dependerán mucho de cada máquina y puesto de trabajo concreto, ya que para algunas máquinas no tendrá ningún sentido porque ya disponen de paneles informativos que proporcionan toda la información, mientras que en otros casos las máquinas pueden ser muy antiguas y no disponen de estos paneles, o simplemente el acceso a la información que proporcionan las máquinas implica invertir una gran cantidad de tiempo. En estos dos últimos casos es donde los encuestados han indicado que podría ser interesante acceder a la información rápidamente con nuevos métodos de comunicación hombre-máquina.

Finalmente, se indicó a los encuestados que realizaran comentarios libres sobre los contenidos evaluados. En la gran mayoría de los casos, se mostró un gran interés por los contenidos 3D, ya sea para formación a través de simuladores de RV, para mantenimiento a través del acceso a los contenidos con dispositivos móviles tradicionales o bien con RA y para acceso a datos del GD, segmentados y con un acceso rápido y fácil.

Podemos concluir por tanto que técnicamente estas tecnologías suponen un impacto positivo en las empresas y sus procesos, siendo necesario encontrar un punto de encuentro entre las ventajas de estos contenidos y el coste de generación de los mismos. A modo de resumen, estos serían los puntos principales:

\begin{itemize}
\item Las empresas deben encontrar los contenidos susceptibles de ser digitalizados con RA y RV, ya sea para formación, mantenimiento o datos IoT.
\item Es importante contar con una plataforma que reduzca el coste considerablemente y facilite la generación de los contenidos
\item Se debe dotar a los empleados no sólo de las herramientas hardware necesarias, sino también de una formación de calidad para evitar rechazo a la tecnología
\item Se debe alinear el uso de estas tecnologías con otras estrategias de digitalización para evitar inversiones duplicadas
\item Los contenidos digitalizados con RA y RV aportan ventajas evidentes, pero el coste de su generación debe ser el adecuado
\end{itemize}

Con el objetivo de facilitar a las empresas la creación de una estrategia eficiente para la implantación de las tecnologías de RA, RV y GD, se ha creado una guía de referencia \footnote{https://www.facyl.es/wp-content/uploads/2021/03/Documento\_Guidelines\_CONECT4.pdf}, que se ha publicado en el apartado de descargas del proyecto CONECT4.

\section{Conclusiones}
Aunque las empresas industriales son conscientes de la necesidad que el mercado les está exigiendo, se encuentran con que, tanto el desconocimiento de las nuevas tecnologías, como su escasa experiencia en este campo, frenan la implantación de tecnologías habilitadoras de la I40. 

El proyecto CONECT4 pretendía generar conocimiento con el fin de sentar las bases sobre las que plantear una nueva metodología de comunicación hombre-máquina con visualización distribuida de datos, para la generación de un ecosistema de aprendizaje orientado al mantenimiento predictivo en la industria. Por otra parte, analiza entornos representativos de los procesos productivos con el fin de evaluar la aplicación de nuevas soluciones digitales sobre diferentes tipologías de empresas en su concepción inicial, proporcionando un documento con guías de referencia para la implantación efectiva de THDs en procesos industriales de mantenimiento para las empresas de Castilla y León. La difusión y publicación de los resultados, a su vez, facilita la visualización de la I40 como algo factible en el entorno empresarial, mostrando a las empresas un nuevo escenario de fábricas digitalizadas o “Smart factories”, capaces de adaptarse a las nuevas necesidades de forma flexible y ajustando los procesos de mantenimiento predictivo.

Una vez finalizado el proyecto, podemos concluir, que se ha alcanzado satisfactoriamente el objetivo general, tras obtener un modelo de ecosistema de aprendizaje validado con trabajadores de las plantas a través de pruebas de concepto desarrolladas con la implementación de THDs de la I40. Estas soluciones se han apoyado en técnicas de visualización aumentada y conocimiento predictivo, donde personas, objetos y sistemas pueden conectarse a través de nuevos métodos de comunicación no intrusivos, logrando mejorar el conocimiento sobre los procesos de mantenimiento industrial de forma más eficiente y predictiva.

\bibliographystyle{unsrtnat}
\bibliography{References}  

\begin{thebibliography}{20}
\providecommand{\natexlab}[1]{#1}
\providecommand{\url}[1]{\texttt{#1}}
\expandafter\ifx\csname urlstyle\endcsname\relax
  \providecommand{\doi}[1]{doi: #1}\else
  \providecommand{\doi}{doi: \begingroup \urlstyle{rm}\Url}\fi

\bibitem[Choi et~al.(2015)Choi, Jung, and Noh]{Choi2015}
Sangsu Choi, Kiwook Jung, and Sang~Do Noh.
\newblock Virtual reality applications in manufacturing industries: Past
  research, present findings, and future directions.
\newblock \emph{Concurrent Engineering}, 23, 03 2015.
\newblock \doi{10.1177/1063293X14568814}.

\bibitem[Alexander et~al.(2016)Alexander, Westhoven, and
  Conradi]{Alexander2016}
Thomas Alexander, Martin Westhoven, and Jessica Conradi.
\newblock Virtual environments for competency-oriented education and training.
\newblock volume 498, pages 23--29, 07 2016.
\newblock ISBN 978-3-319-42069-1.
\newblock \doi{10.1007/978-3-319-42070-7_3}.

\bibitem[Bacca-Acosta et~al.(2014)Bacca-Acosta, Baldiris, Fabregat, Graf, and
  Kinshuk]{Bacca2014}
Jorge Bacca-Acosta, Silvia Baldiris, Ramón Fabregat, Sabine Graf, and
  Dr~Kinshuk.
\newblock Augmented reality trends in education: A systematic review of
  research and applications.
\newblock \emph{Educational Technology and Society}, 17:\penalty0 133--149, 10
  2014.

\bibitem[Di~Serio et~al.(2013)Di~Serio, Ibáñez, and
  Delgado-Kloos]{DiSerio2013}
Angela Di~Serio, María Ibáñez, and Carlos Delgado-Kloos.
\newblock Impact of an augmented reality system on students' motivation for a
  visual art course.
\newblock \emph{Computers \& Education}, 68:\penalty0 586–596, 10 2013.
\newblock \doi{10.1016/j.compedu.2012.03.002}.

\bibitem[Tao et~al.(2019)Tao, Zhang, Liu, and Nee]{Tao2019}
Fei Tao, He~Zhang, Ang Liu, and A.~Y.~C. Nee.
\newblock Digital twin in industry: State-of-the-art.
\newblock \emph{IEEE Transactions on Industrial Informatics}, 15\penalty0
  (4):\penalty0 2405--2415, 2019.
\newblock \doi{10.1109/TII.2018.2873186}.

\bibitem[O'Donovan et~al.(2016)O'Donovan, Bruton, and
  O'~Sullivan]{Odonovan2016}
Peter O'Donovan, Ken Bruton, and Dominic O'~Sullivan.
\newblock Case study: The implementation of a data-driven industrial analytics
  methodology and platform for smart manufacturing.
\newblock \emph{International Journal of Prognostics and Health Management}, 7,
  10 2016.
\newblock \doi{10.36001/ijphm.2016.v7i3.2414}.

\bibitem[Macchi et~al.(2018)Macchi, Roda, Negri, and Fumagalli]{Macchi2018}
Marco Macchi, Irene Roda, Elisa Negri, and Luca Fumagalli.
\newblock Exploring the role of digital twin for asset lifecycle management.
\newblock \emph{IFAC-PapersOnLine}, 51:\penalty0 790--795, 01 2018.
\newblock \doi{10.1016/j.ifacol.2018.08.415}.

\bibitem[Erikstad(2017)]{Erikstad2017}
Stein Erikstad.
\newblock Merging physics, big data analytics and simulation for the
  next-generation digital twins.
\newblock 09 2017.

\bibitem[Banerjee et~al.(2017)Banerjee, Mittal, Dalal, and Joshi]{Banerjee2017}
Agniva Banerjee, Sudip Mittal, Raka Dalal, and Karuna Joshi.
\newblock Generating digital twin models using knowledge graphs for industrial
  production lines.
\newblock 06 2017.
\newblock \doi{10.1145/3091478.3162383}.

\bibitem[Weyer et~al.(2015)Weyer, Schmitt, Ohmer, and Gorecky]{Weyer2015}
Stephan Weyer, Mathias Schmitt, Moritz Ohmer, and Dominic Gorecky.
\newblock Towards industry 4.0 - standardization as the crucial challenge for
  highly modular, multi-vendor production systems.
\newblock volume~48, pages 579--584, 12 2015.
\newblock \doi{10.1016/j.ifacol.2015.06.143}.

\bibitem[Grieves(2015)]{Grieves2015}
Michael Grieves.
\newblock Digital twin: Manufacturing excellence through virtual factory
  replication.
\newblock 03 2015.

\bibitem[Tao et~al.(2018{\natexlab{a}})Tao, Sui, Liu, Qi, Zhang, Song, Guo, Lu,
  and Nee]{Tao2018}
Fei Tao, Fangyuan Sui, Ang Liu, Qinglin Qi, Meng Zhang, Boyang Song, Zirong
  Guo, Stephen Lu, and Andrew Nee.
\newblock Digital twin-driven product design framework.
\newblock \emph{International Journal of Production Research}, 57:\penalty0
  1--19, 02 2018{\natexlab{a}}.
\newblock \doi{10.1080/00207543.2018.1443229}.

\bibitem[Uhlemann et~al.(2017)Uhlemann, Lehmann, and
  Steinhilper]{UHLEMANN2017335}
Thomas H.-J. Uhlemann, Christian Lehmann, and Rolf Steinhilper.
\newblock The digital twin: Realizing the cyber-physical production system for
  industry 4.0.
\newblock \emph{Procedia CIRP}, 61:\penalty0 335--340, 2017.
\newblock ISSN 2212-8271.
\newblock \doi{https://doi.org/10.1016/j.procir.2016.11.152}.
\newblock URL
  \url{https://www.sciencedirect.com/science/article/pii/S2212827116313129}.
\newblock The 24th CIRP Conference on Life Cycle Engineering.

\bibitem[Qi et~al.(2018)Qi, Tao, Zuo, and Zhao]{QI2018237}
Qinglin Qi, Fei Tao, Ying Zuo, and Dongming Zhao.
\newblock Digital twin service towards smart manufacturing.
\newblock \emph{Procedia CIRP}, 72:\penalty0 237--242, 2018.
\newblock ISSN 2212-8271.
\newblock \doi{https://doi.org/10.1016/j.procir.2018.03.103}.
\newblock URL
  \url{https://www.sciencedirect.com/science/article/pii/S2212827118302580}.
\newblock 51st CIRP Conference on Manufacturing Systems.

\bibitem[Glass et~al.(2018)Glass, Meißner, Gebauer, Stürmer, and
  Metternich]{Glass2018}
Rupert Glass, Alyssa Meißner, Christopher Gebauer, Sandra Stürmer, and
  Joachim Metternich.
\newblock Identifying the barriers to industrie 4.0.
\newblock \emph{Procedia CIRP}, 72:\penalty0 985--988, 01 2018.
\newblock \doi{10.1016/j.procir.2018.03.187}.

\bibitem[Chesworth(2018)]{Chesworth2018}
Daren Chesworth.
\newblock Industry 4.0 techniques as a maintenance strategy (a review paper).
\newblock 01 2018.
\newblock \doi{10.13140/RG.2.2.18116.32644}.

\bibitem[Helu et~al.(2017)Helu, Hedberg, and Barnard~Feeney]{Helu2017}
Moneer Helu, Thomas Hedberg, and Allison Barnard~Feeney.
\newblock Reference architecture to integrate heterogeneous manufacturing
  systems for the digital thread.
\newblock \emph{CIRP Journal of Manufacturing Science and Technology}, 05 2017.
\newblock \doi{10.1016/j.cirpj.2017.04.002}.

\bibitem[Tao et~al.(2018{\natexlab{b}})Tao, Cheng, Qi, Zhang, Zhang, and
  Sui]{Tao2018b}
Fei Tao, Jiangfeng Cheng, Qinglin Qi, Meng Zhang, He~Zhang, and Fangyuan Sui.
\newblock Digital twin-driven product design, manufacturing and service with
  big data.
\newblock \emph{The International Journal of Advanced Manufacturing
  Technology}, 94, 02 2018{\natexlab{b}}.
\newblock \doi{10.1007/s00170-017-0233-1}.

\bibitem[Carlsson(2018)]{Carlsson2018}
Christer Carlsson.
\newblock Decision analytics mobilized with digital coaching.
\newblock \emph{Intelligent Systems in Accounting, Finance and Management},
  25:\penalty0 3--17, 01 2018.
\newblock \doi{10.1002/isaf.1421}.

\bibitem[Gento~Municio et~al.(2020)Gento~Municio, Pimentel, and
  Pascual]{Gento2020}
Ángel Gento~Municio, Carina Pimentel, and José Pascual.
\newblock Lean school: an example of industry-university collaboration.
\newblock \emph{Production Planning \& Control}, 32:\penalty0 1--16, 03 2020.
\newblock \doi{10.1080/09537287.2020.1742373}.

\end{thebibliography}

\end{document}